# Magnetoelectric effect in antiferromagnetic multiferroic Pb(Fe$_{1/2}$Nb$_{1/2}$)O$_3$ and its solid solutions with PbTiO$_3$

V. V. Laguta[1,2], V. A. Stephanovich[2], I. P. Raevski[3], S.I. Raevskaya[3], V.V. Titov[3], V.G. Smotrakov[3], V.V. Eremkin[3]

[1]Institute of Physics AS CR, Cukrovarnicka 10, 162 53 Prague, Czech Republic

[2]*Institute of Physics, Opole University, Oleska 48, 45-052 Opole, Poland*

[3]Research Institute of Physics and Physical Faculty, Southern Federal University, Stachki Ave. 194, 344090, Rostov-on-Don, Russia.

**Abstract**

Antiferromagnets (AFMs) are presently considered as promising materials for applications in spintronics and random access memories due to the robustness of information stored in AFM state against perturbing magnetic fields (P. Wadley et al., Science **351**, 587 (2016)). In this respect, AFM multiferroics maybe attractive alternatives for conventional AFMs as the coupling of magnetism with ferroelectricity (magnetoelectric effect) offers an elegant possibility of electric field control and switching of AFM domains. Here we report the results of comprehensive experimental and theoretical investigations of the quadratic magnetoelectric (ME) effect in single crystals and high-resistive ceramics of Pb(Fe$_{1/2}$Nb$_{1/2}$)O$_3$ (PFN) and (1- x)Pb(Fe$_{1/2}$Nb$_{1/2}$)O$_3$−xPbTiO$_3$ (PFN−xPT). We are interested primarily in the temperature range of multiferroic phase, $T < 150$ K, where the ME coupling coefficient is extremely large (as compared to well-known multiferroic BiFeO$_3$) and shows sign reversal at paramagnetic-to-antiferromagnetic phase transition. Moreover, we observe strong ME response nonlinearity in the AFM phase in the magnetic fields of only few kOe. To describe the temperature and magnetic field dependencies of the above unusual features of ME effect in PFN and PFN-xPT, we use simple phenomenological Landau approach which explains experimental data surprisingly well. Our ME measurements demonstrate that the electric field of only 20-25 kV/cm is able to switch the AFM domains and align them with ferroelectric ones even in PFN ceramic samples.

## I. Introduction

Pb(Fe$_{1/2}$Nb$_{1/2}$)O$_3$ (PFN) and its solid solutions with PbTiO$_3$ (PT) and PbZr$_{1-x}$Ti$_x$O$_3$ (PZT) are among the most intensively studied multiferroic materials, see recent publications [1,2,3,4,5]. The disorder in PFN perovskite structure is due to the presence of two different ions, magnetic Fe$^{3+}$ and non-magnetic Nb$^{5+}$, at the same octahedral position. Nevertheless, it undergoes quite normal ferroelectric phase transition at the temperature T$_C$≈370 K with large enough remanent polarization (see, e.g., Refs. [6,7]). On cooling, PFN undergoes antiferromagnetic (AFM) phase transition at the Neel temperature T$_N$≈150 K and finally at low temperatures (11-12 K) it freezes into a spin glass state which, however, coexists with the above mentioned long-range ordered AFM phase. The latter coexistence has recently been studied in details both experimentally and theoretically [3,8].

Magnetoelectric (ME) effect had been observed in PFN many years ago, starting from $1980^{th}$. To the best of our knowledge, this have been done only for single crystals and in the AFM phase [1,9]. We have found recently [10], that sizeable ME effect exists also in the paramagnetic (PM) phase of both PFN single crystals and ceramics as well as in PFN-PT solid solution ceramics. This ME effect disappears only in the paraelectric phase, where there is no spontaneous polarization. It is worth noting that while the linear ME effect in PFN is well researched [9], the quadratic ME coupling is much less studied despite its huge value ($\beta_{333}=10^{-17} – 10^{-16}$ s/A [1,11]), which is almost three orders of magnitude larger than that in the well-known AFM multiferroic $BiFeO_3$ ($\beta_{333}=2.1 \cdot 10^{-19}$ s/A [12,13]). The reason for this difference is still not clarified.

Large ME coupling in PFN, especially in ceramic samples may be attractive for applications in ME memory elements and spintronics as AFM domains are almost unsusceptible to external magnetic fields which preserves well the stored information. The coupling between ferroelectricity and antiferromagnetism in PFN offers intriguing possibility of electric filed control and switching of AFM domains. Such electric field switching of AFM domains has been demonstrated in $BiFeO_3$ single crystals [14], but to the best of our knowledge it was never observed in ceramics.

Another important question which we want to clarify in this paper is the behavior of the ME coupling between disordered (dynamically or statically) spin ensemble and electric polarization in magnetoelectrics with spin-glass or superparamagnetic phases. In this context, the PFN diluted by PT, is an almost ideal system as both ferroelectric and magnetic phases can be predictably modified by varying the Ti content in PFN−xPT solid solution [**Error! Bookmark not defined.**,**Error! Bookmark not defined.**,7].

Interestingly, the phenomenological Landau theory predicts the ME coupling increase on cooling proportional to the square of magnetic susceptibility [10,15]. Such an increase has not been experimentally observed in PFN [1], though its susceptibility increases substantially at temperatures below $T_N \approx 150$ K [2]. On the contrary, ME coupling was reported to decrease below the spin-glass freezing temperature $T_g \approx 11$ K [1].

In the present paper, we report the results of comprehensive investigations of the quadratic ME effect in crystals and high-resistive ceramics of $Pb(Fe_{1/2}Nb_{1/2})O_3$ and $(1-x)Pb(Fe_{1/2}Nb_{1/2})O_3–xPbTiO_3$. Usage of ceramic samples enables performing dielectric and ME response measurements up to the temperatures 400-450 K without marked influence of conductivity. The main attention is payed to low ($T$<150 K) temperatures, where the ME coupling coefficient is extremely large as compared, for instance, to $BiFeO_3$ and shows sign reversal at paramagnetic-to-antiferromagnetic phase transition. Moreover, we observed strong ME response nonlinearity in the AFM phase in the fields of only a few kOe. We also present the results of ME measurements in PFN ceramics with $90^0$ and $180^0$ switching of electric polarization in the AFM phase which demonstrate that the alignment of electric domains leads to corresponding alignment of magnetic domains. To describe the temperature and magnetic field dependencies of the ME effect we use a simple Landau theory of phase transitions which explains experimental data surprisingly well.

The plan of our paper is the following. After a short description of the applied experimental methods (Sec. II), we report on our experimental exploration of the ME effect firstly in pure PFN (both single crystals and ceramics, Sec. IIIA) and then in PFN-PT and PFN-PZT solid solutions (Sec. IIIB). Sec. IV presents phenomenological theory of quadratic ME coupling in both paramagnetic and AFM phases. Finally, in Secs. V and VI, we discuss the results obtained and make conclusions.

## II. Experimental

Single crystals of PFN–xPT at x=0, 0.03, and 0.2 were grown by the spontaneous crystallization from the PbO - $B_2O_3$ flux in the temperature range from 1010 down to 850 $^0$C (see Refs. [2,16] for details). The fabricated crystals were cubic-shaped with the edges up to 4-6 mm and the faces parallel to (100) planes of the prototype perovskite structure. Chemical composition of the crystals obtained has been determined by the electron probe X-ray microanalyzer "Camebax- Micro", using PFN and $PbTiO_3$ as reference samples. In our measurements, crystals were cut either along (111) or (100) planes depending on spontaneous polarization direction.

Ceramic samples of PFN–xPT (x=0, 0.05) solid solution have been fabricated by solid-state reaction route using high-purity $Fe_2O_3$, $Nb_2O_5$, PbO, and $TiO_2$. These oxides were batched in stoichiometric proportions and 1 wt. % of $Li_2CO_3$ was added to the batch. This addition promotes formation of the PFN perovskite modification and inhibits conductivity [17]. The sintering has been performed at 1030-1070 $^0$C for 2 hours in a closed alumina crucible. The density of the obtained ceramics was about 92-97 % of theoretical one. X-ray diffraction analysis showed that all investigated compositions were single-phased and had a perovskite-type structure.

Typical sample size for ME measurements was 2.5x5x0.9 $mm^3$. The electrodes for measurements were deposited by silver paint (SPI Supplies, USA). Before the ME measurements, ceramic samples were poled at room temperature by *dc* electric field of 10 kV/cm for 10 min. Single crystals were poled at *T*=77-180 K to avoid influence of conductivity which makes impossible crystals poling at higher temperatures. Poled samples were tested by measuring the piezoelectric coefficient, $d_{31}$, by the standard resonance - antiresonance method.

The ME coefficient was determined by a dynamic method [18] as a function of bias magnetic field *H* at small ac field *h* = 1-5 Oe and frequencies 0.2-1 kHz (at this low frequencies ME response did not depend on frequency) by measuring the voltage across the sample utilizing a lock-in-amplifier with high impedance preamplifier. *ac* and *dc* magnetic fields were applied normally to the surface of the sample with electrodes. In every ME experiment, more than two runs were repeated with the direction of *H* reversed and the change of the signal sign was confirmed. In this way, a possible spurious signal was segregated from a true ME one whose sign is dependent on the *PH* product.

In our experiment, the ME effect is manifested as a polarization $P$ induced by a small ac magnetic field $h_{ac}$ under application of dc field $H_{dc}$ [18,19]. The magnetic field induced components of the polarization can be obtained from the free energy expansion [20]:

$$F(\bar{E}, \bar{H}) = F_0 - P_i^s E_i - M_i^s H_i - \frac{1}{2}\varepsilon_0 \varepsilon_{ij} E_i E_j - \frac{1}{2}\mu_0 \mu_{ij} H_i H_j - \alpha_{ij} E_i H_j - \frac{1}{2}\beta_{ijk} E_i H_j H_k - ... \quad (1a)$$

$$P_i = -\frac{\partial F}{\partial E_i} = P_i^s + \varepsilon_0 \varepsilon_{ij} E_j + \alpha_{ij} H_j + \frac{1}{2}\beta_{ijk} H_j H_k + ... \quad (1b)$$

where $i,j,k=x,y,z$ are Cartesian coordinates, $P^s$ and $M^s$ are, respectively, the spontaneous polarization and magnetization; $\mu_{ij}$ and $\varepsilon_{ij}$ are, respectively, magnetic and dielectric permittivities ($\mu_0$ and $\varepsilon_0$ are the vacuum permittivities in SI units); $\alpha_{ij}$ and $\beta_{ijk}$ are linear and linear-quadratic ME coupling coefficients, respectively. Using collinear *dc* and *ac* magnetic fields $H = H_{dc} + h\sin\omega t$, the first harmonic amplitude of the *ac* polarization detected by lock-in detector is:

$$P_i(T) = \alpha_{ij}(T) h_j + \beta_{ijj}(T) H_j h_j. \quad (1c)$$

Since the linear ME coupling in PFN is much smaller than the quadratic one and we are interested primarily in the quadratic ME effects, we neglect the α term hereafter. Technically, the quadratic ME effect is described by a third rank tensor $\beta_{ijk}$ for the arbitrary orientations of polarization and magnetic field vectors. In our experimental geometry, when both magnetic field and polarization are aligned along one crystallographic direction ([111] or [100] depending on spontaneous polarization direction), the measured ME coefficient is $\beta_{333}$. Obviously, only an effective coupling constant $\beta_{333}$, which represents an average of the different elements of the $\beta_{ijk}$ tensor, should be considered in ceramics. For simplification, we will omit indexes in the $\beta_{333}$ element. Note that a microscopic theory of the PME effect for $C_{3h}$ symmetry had been presented in Ref. [21].

The ME response was measured as the voltage induced by the *ac* polarization in a sample. The ME voltage is determined from Eq. (1b) as

$$U_{ME} = \frac{dP_{ac}}{dt}(\omega C + 1/R_i)^{-1} \approx \frac{\beta H_{dc} hS}{C}, \quad (2)$$

where $C$ and S are, respectively, the sample capacitance and area. The expression for $U_{ME}$ is valid under the condition $(\omega C)^{-1} \ll R_i$, where $R_i \sim 10^9$ Ohm is the impedance of lock-in-amplifier with preamplifier. This relation is always fulfilled at the frequencies 0.2-1 kHz due to high samples' capacitance.

## III. Experimental results

As the dielectric and magnetic properties of our PFN single crystals and ceramics have been studied previously (see, e.g. Refs. [2,6,7]) here we present the results of ME effect measurements only. Also, here we primarily use the PFN–xPT solid solution crystals because ceramic samples contained ferromagnetic or superparamagnetic impurity phases, which masked intrinsic ME response at low magnetic fields.

### A. ME effect in PFN crystals and ceramics

Fig. 1a shows temperature dependence of ME voltage measured in PFN crystal when magnetic field is parallel to spontaneous polarization, aligned along the [111] crystal direction. One can see that ME response strongly increases on cooling below the Neel temperature and changes the sign at the transition from PM to AFM phase. Note that similar effect has been observed previously (in measurements of electric field induced ME moment) but with lower resolution [1]. The sign change of ME voltage is well seen in highly-resistive PFN ceramics (Fig. 1b). In ceramic samples, ME signal can be measured up to the temperature of FE phase transition as it was demonstrated in Ref. [10**Error! Bookmark not defined.**]. Note that the quadratic ME coefficient in AFM phase of PFN ($\beta_{333}$=2.5·10$^{-17}$ s/A in a crystal at T=10 K and $\beta$=1.0·10$^{-16}$ s/A in ceramics at T=20 K) is 2-3 orders of magnitude higher than that in BiFeO$_3$ crystal ($\beta_{333}$=2.1·10$^{-19}$ s/A at $T$=4.2 K [12]). Such a large ME coefficient in ceramics indicates that sublattice magnetization is preferably aligned along electric polarization so that magnetization can in principle be switched by an electric field in AFM phase.

These data also suggest that the paramagnetoelectric (PME) contribution is non-zero in magnetically ordered phase. It competes with the ME contribution related to AFM order parameter which has the opposite sign. This is well seen in ceramics, where the inversion point is shifted to about 100 K due to lower Neel temperature typical of Li-doped PFN ceramics [22]. These two contributions to the ME response in the free energy expansion have the following form [10**Error! Bookmark not defined.**] (see also below Sec. IV):

$$G_{ME} = \frac{1}{2}\left(\xi_{MP}M^2 + \xi_{LP}L^2\right)P^2. \qquad (3)$$

Here $L$ is the AFM order parameter and $\xi_{MP}$ and $\xi_{LP}$ are biquadratic ME coefficients that couple corresponding order parameters.

Essentially non-linear dependence of ME response on applied field in the AFM phase (Fig. 2a) may be generated by the competition between two terms in Eq. (3) as they have the opposite signs and different temperature and field dependencies. In particular, the (positive) AFM order parameter saturates at low temperatures, while the absolute value of the (negative) second term still increases as square of susceptibility. However, this is only one of the possible reasons for the observed non-linearity. The second mechanism of the ME effect non-linearity can be caused by spin-flip and spin-flop transitions. But for PFN the critical fields of these transitions are much higher than those used in our experiment. Namely, the estimations of both above fields from exchange energies [**Error! Bookmark not defined.**,23] show that they are larger than 50 and 250 kOe, respectively. Magnetostriction may also play a role in the AFM phase. However, as it will be shown in Section IV, even simple phenomenological Landau – Ginzburg - Devonshire approach with appropriate coefficients in the free energy expansion allows one to explain satisfactorily the non-linearity of ME response.

In the PM phase where $L=0$, the ME voltage becomes a linear function of magnetic field (Fig. 2b) as only PM term contributes to the ME effect now. The anomaly at zero magnetic fields is produced by parasitic superparamagnetic phase in ceramics.

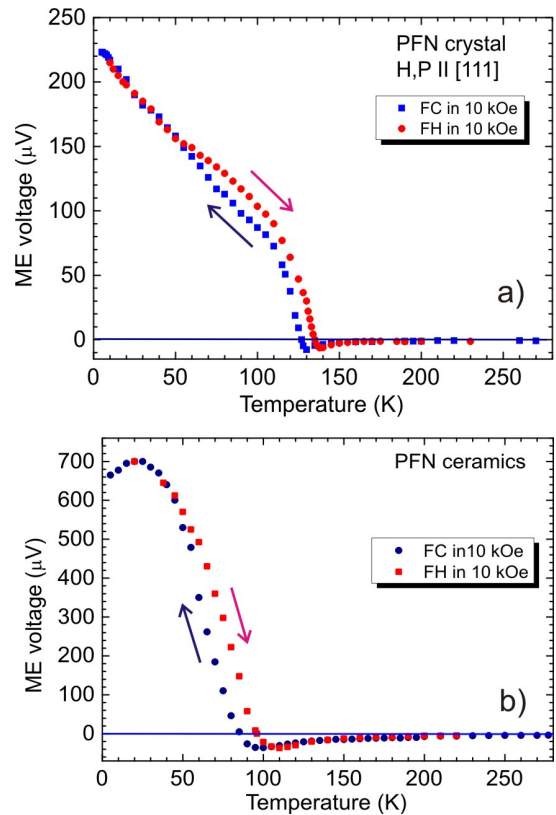

Fig. 1. Temperature dependence of ME voltage measured in poled PFN crystal (a) and ceramics (b) under the field of 10 kOe. The magnetic field *H* is applied parallel to the electric polarization *P*.

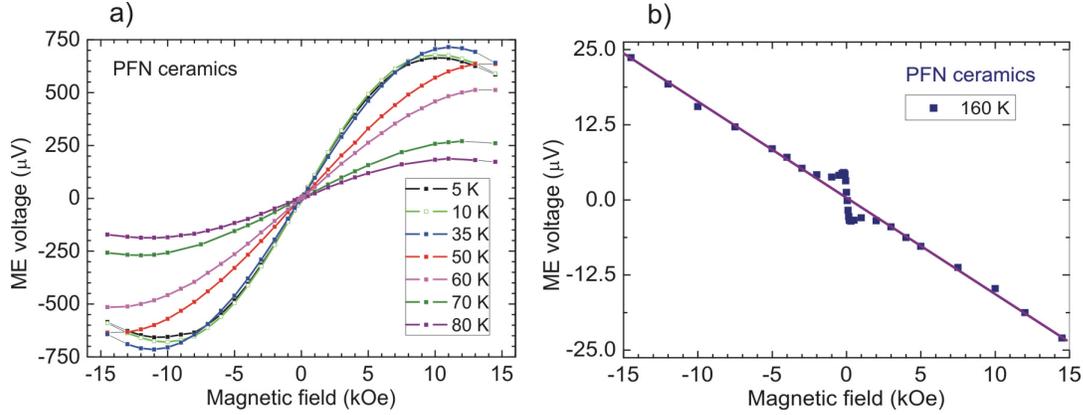

Fig. 2. ME voltage versus magnetic field in AFM (a) and PM (b) phases of PFN ceramics at selected temperatures shown in the legends. The anomaly around zero magnetic field in the panel (b) is due to parasitic superparamagnetic phase. Solid lines show the fits to Eqs. (33), (34) of section IV.

**B. ME effect in PFN-xPT solid solution**

An addition of PT to PFN permits to manipulate the parameters of magnetic and FE phase transitions in the PFN-xPT solid solution. For instance at the PT concentration $x$ larger than about 10%, the long-range AFM ordering of $Fe^{3+}$ spins is suppressed [2] and the ferroelectric phase has tetragonal symmetry [4,7]. For the reader's convenience, Fig. 3 reports the magnetic and electric phase diagrams of PFN-xPT solid solution plotted on the base of published experimental data [2,4,7].

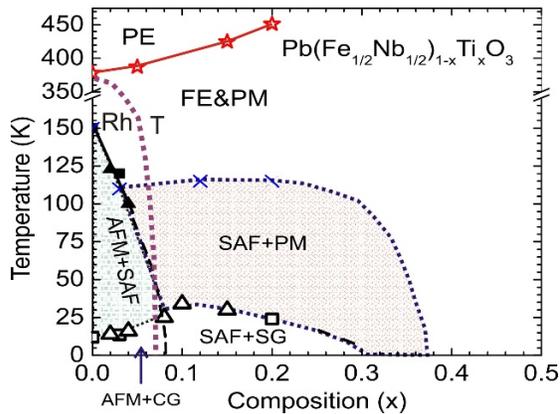

Fig. 3. Phase diagrams of PFN−xPT solid solution. Solid and dashed lines separate different magnetic and electric phases. The red line with stars separates paraelectric (PE) and ferroelectric (FE) phases. The dotted brown line is the morphotropic phase boundary between the rhombohedral (Rh) and tetragonal (T) phases. SAF indicates the superantiferromagnetic state. SG and CG stand for spin glass and spin cluster glass states, respectively [1,**Error! Bookmark not defined.**].

Fig. 4a shows temperature dependence of the ME voltage measured in the PFN-0.03PT crystal. The weak dilution of PFN by non-magnetic Ti ions lowers the Neel temperature from 150 K to 120 K but does not change essentially the temperature behavior of the ME response. The latter becomes much more non-linear *vs* the magnetic field as compared to undoped material (Fig. 4b). Both Fig. 4a and Fig. 4b show some anomalous variation in ME response at the temperatures 30-40 K. For instance, the magnetic field dependence of ME voltage in Fig 4b changes its slope from negative to positive at H>13 kOe. The physical origin of this feature is presently not clear.

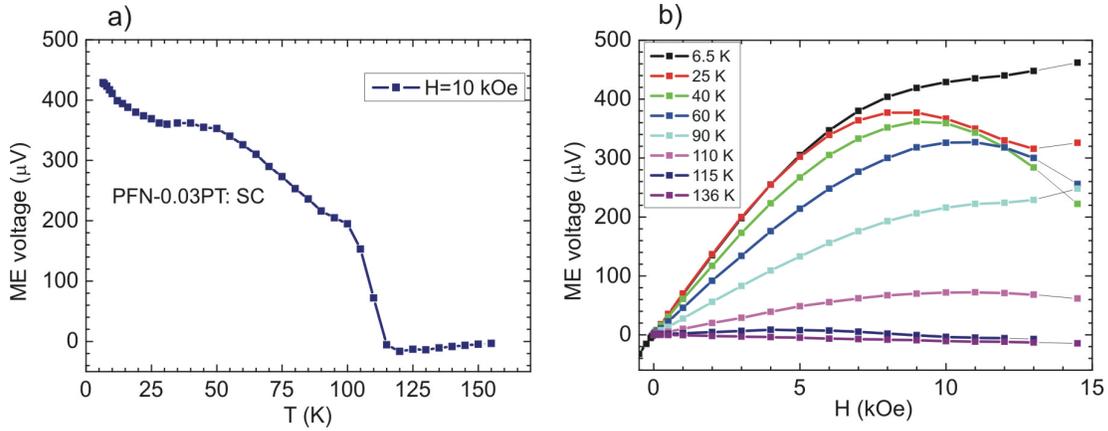

Fig. 4. Temperature (a) and magnetic field (b) dependencies of ME voltage in PFN−0.03PT solid solution crystal under magnetic field parallel to the [111] crystal direction.

It is expected that further increase of the Ti concentration would weaken the AFM phase so that the negative PM-like (i.e. PME) contribution may become dominating. Indeed, the negative PME contribution dominates at all the temperatures already for PFN-0.05PT (Fig. 5a) as its Neel temperature is essentially shifted to low temperatures around 55-60 K. Besides, as it can be seen from the phase diagram (Fig. 3), the magnetic state of this composition is complex enough. It contains superposition of various phases: PM, superantiferromagnetic (SAF), AFM and spin glass (SG). Therefore, a complex interplay between electric and different magnetic order parameters occurs in PFN-0.05PT. As a result, the ME response being a perfect linear function of magnetic field in the PM phase becomes strongly nonlinear at lower temperatures in the magnetically ordered phase even at low fields of a few kOe (Fig. 5b). This composition is also close to the morphotropic boundary between the tetragonal and rhombohedral phases. We cannot thus exclude the rotation of the polarization from the [111] rhombohedral axis to the [100] tetragonal one on cooling that will essentially influence the ME response. At $T \approx 35$ K, the ME coefficient is negative with large enough modulus, $\beta = -1.5 \cdot 10^{-17}$ s/A. We calculate this value from the slope of $U_{ME}(H)$ curve at low fields. However, ME coefficient decreases almost to zero on further cooling even at low fields due to interplay between the AFM phase and spin glass state.

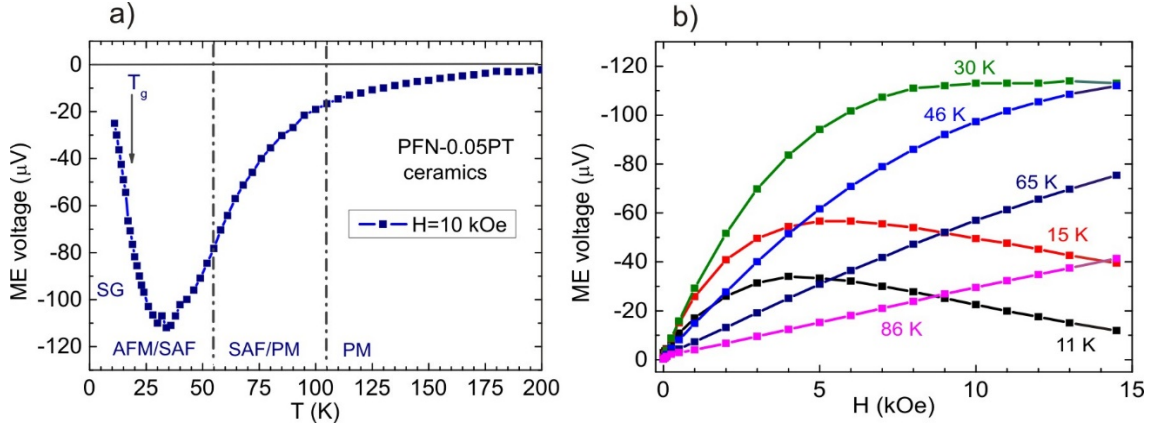

Fig. 5. Temperature (a) and magnetic field (b) dependencies of ME voltage in PFN−0.05PT solid solution ceramics. Vertical dashed lines in panel (a) separate different magnetic phases.

Further dilution of PFN by Ti (PFN−0.2PT composition) suppresses the long-range ordering of $Fe^{3+}$ spins so that only a spin-glass state emerges below the freezing temperature about 23 K. This can be seen from zero field cooled (ZFC) and field cooled (FC) magnetic susceptibilities shown in the inset of Fig. 6a and phase diagram in Fig. 3. The ME voltage thus increases on cooling according to temperature variation of the FC magnetic susceptibility as polarization and dielectric susceptibility are almost constant at these temperatures, see Fig. 6a. The ME response is negative, PME-like, but large enough at low temperatures ($\beta = -1.5 \cdot 10^{-17}$ s/A at $T$=10 K), being comparable with that in the AFM phase of PFN. The ME response is a perfect linear function of applied magnetic field even at the lowest temperatures, Fig. 6b. The data presented in Fig. 6 can be thus described by the following relation derived by us in Ref. [**Error! Bookmark not defined.**10]:

$$\beta(T) = -P_S(T)\chi_{FE}(T)(\chi_M(T))^2 \xi_{MP} . \qquad (4)$$

The solid line in Fig. 6a is a fit to Eq. (4) assuming that the only temperature dependent quantity is the magnetic susceptibility shown in the inset of Fig. 6a. Taking into account that FC susceptibility in SAF and SG phases depends on magnetic field (it was measured at 500 Oe, while ME voltage was measured at 10 kOe), we obtain the qualitatively good coincidence with experiment.

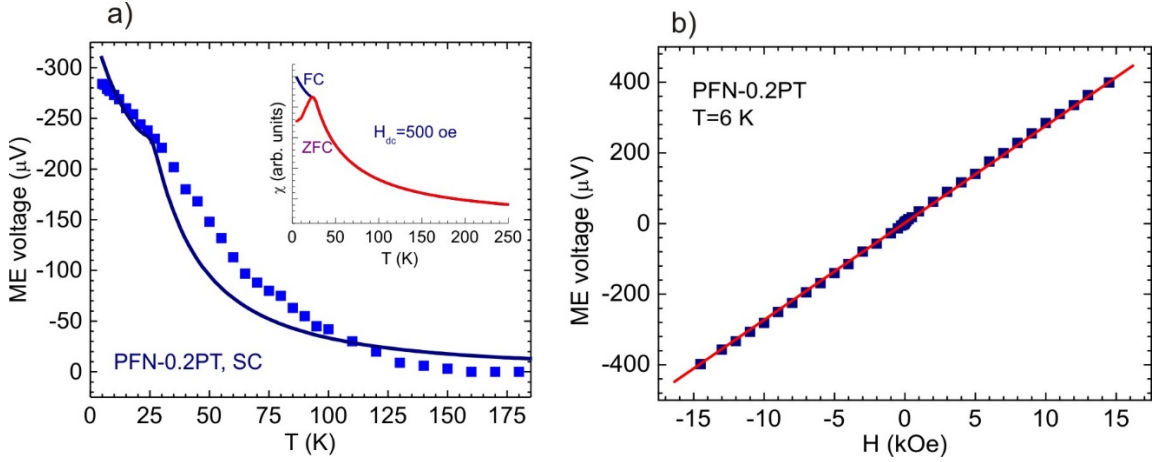

Fig. 6. (a) Temperature dependence of ME voltage in PFN−0.2PT solid solution crystal measured under the field 10 kOe. Solid line is a fit to Eq. (3) assuming temperature dependence of only magnetic susceptibility presented in the inset. (b) Magnetic field dependence of ME voltage at T=6 K.

## IV. Magnetoelectric coefficient in the simple phenomenological model
### A. Statement of the problem and free energy function

The electric polarization of a magnetoelectric depends on magnetic field $H$, $P_{ME} \equiv P(H)$. Below we will find $P(H)$ from the minimum of corresponding free energy, but for now we suppose that the function $P(H)$ is known. The magnetic field in experiment consists of two parts

$$H = H_{dc} + h_{ac}, h_{ac} = h\sin\omega t, h_{ac} \ll H_{dc}.$$

As $h_{ac} \ll H_{dc}$, we can expand the polarization $P(H) = P(H_{dc} + h_{ac})$ in power series in small $h_{ac}$. This yields

$$P(H_{dc} + h_{ac}) \approx P(H_{dc}) + h_{ac}\left.\frac{dP}{dH}\right|_{H=H_{dc}} + ...$$

Now, according to Eq. (1b), $P(H_{dc})$ is proportional to $H_{dc}^2$, i.e. rewriting Eq. (1b) in scalar notation, we obtain

$$P(H_{dc}) = P_0 + \frac{1}{2}\beta H_{dc}^2,$$

where $P_0$ comprises three first terms in Eq. (1b) and $\beta$ is a quadratic ME coupling coefficient. The term $P(H_{dc})$ is inaccessible by experiment with lock-in detection on the frequency of $h_{ac}$ field so that we have for experimentally measurable polarization

$$P_{ac}(H) \approx h\left.\frac{dP}{dH}\right|_{H=H_{dc}}.$$

As measured ME voltage is related to the polarization as $U_{ME} = (S/C)P_{ac}(H)$, where S is an area of electrode and C is a sample capacity, we finally obtain

$$U_{ME} = \frac{S}{C} P_{ac}(H) \approx \frac{Sh}{C} \frac{dP}{dH}\bigg|_{H=H_{dc}}.$$

As the coefficient $Sh/C$ does not depend on $H = H_{dc}$, i.e. it is constant, our result for $U_{ME}$ will be proportional to $dP/dH$. By comparing Eq. (2) with the last expression, quadratic ME coefficient can be expressed as

$$\beta = \frac{1}{H_{dc}} \frac{dP}{dH}, H = H_{dc}. \tag{5}$$

The aim of the present consideration is to study the magnetic field and temperature dependence of the magnetoelectric coefficient β of Pb(Fe$_{1/2}$Nb$_{1/2}$)O$_3$ to describe the experiment in its PM and AFM phases. Below we suggest simple phenomenological approach which can be used not only for description of ME effect in PFN, but in other AFM multiferroics. For this purpose we use the simplest possible free energy function [**Error! Bookmark not defined.**]:

$$G = G_P + G_M + G_{ME}, G_P = \frac{1}{2}\alpha_P(T)P^2 + \frac{1}{4}\beta_P P^4 - PE, G_{ME} = \frac{1}{2}(\xi_{MP}M^2 + \xi_{LP}L^2)P^2,$$

$$G_M = \frac{1}{2}\alpha_L(T)L^2 + \frac{1}{4}\beta_L L^4 + \frac{1}{2}\alpha_M(T)M^2 + \frac{1}{4}\beta_M M^4 + \frac{1}{2}\xi_{LM}L^2M^2 - MH, \tag{6}$$

where *P*, *M* and *L* are, respectively, ferroelectric (polarization), ferromagnetic (magnetization) and AFM order parameters, *E* and *H* are the strengths of external electric and magnetic fields. The only temperature dependent coefficients are those, where explicit temperature dependence is indicated, namely

$$\alpha_P(T) = \alpha_{P_0}(T - T_C), \alpha_M(T) = \alpha_{M_0}(T - \theta), \alpha_L(T) = \alpha_{L_0}(T - T_N), \tag{7}$$

where T$_C$≈370 K is the temperature of paraelectric - ferroelectric phase transition, T$_N$≈150 K is the Neel temperature and θ≈−520 K is the temperature of (virtual) ferromagnetic phase transition [2]. Below we will use this phase transition temperatures hierarchy to describe the experimentally observed dependencies *β(T,H)*. Note that for our purposes it is sufficient to consider the ferroelectric phase transition of the second kind, i.e. we truncate power series in *G$_P$* on *P$^4$* term assuming that *β$_P$*>0.

**C. Calculation of temperature and magnetic field dependence of magnetoelectric coefficient**

The idea behind this calculation is natural and simple. First, minimizing the free energy (6), we find the field (magnetic and electric) and temperature dependences of order parameters *P*, *M* and *L* and then, taking the derivative *dP/dH* (with respect to the field and temperature dependences of the rest order parameters), obtain the desired dependence *β(T,H)* in paramagnetic (*L*=0, *T>T$_N$*) and antiferromagnetic $(L \neq 0, T < T_N)$ phases of PFN. Note that according to Eq. (5) the quadratic ME coefficient can be expressed as $\beta = 1/H \times dP/dH$. Therefore, the experimentally measured ME voltage is directly proportional to the *dP/dH*.

The equations for equilibrium order parameters assume the form

$$\frac{\partial G}{\partial P} = 0, \frac{\partial G}{\partial M} = 0, \frac{\partial G}{\partial L} = 0, \tag{8}$$

$$\frac{\partial G}{\partial P} = P(\alpha_P + \xi_{MP}M^2 + \xi_{LP}L^2) + \beta_P P^3 - E = 0, \tag{9}$$

$$\frac{\partial G}{\partial M} = M(\alpha_M + \xi_{LM}L^2 + \xi_{MP}P^2) + \beta_M M^3 - H = 0, \tag{10}$$

$$\frac{\partial G}{\partial L} = L(\alpha_L + \xi_{LM}M^2 + \xi_{LP}P^2) + \beta_L L^2. \tag{11}$$

We pay attention here that as parameters *P* and *M* couple with external fields (there are terms *PE* and *MH* in the free energy (6) and hence *E* and *H* in Eqs. (9) and (10)), while *L* does not, this permits to factor the equation (11) for *L* considering separately the cases *L*=0 (PM phase) and $L \neq 0$ (AFM phase). This eventually yields the nonlinear dependence *β(H)* in AFM phase.

Actually, the set of equations (8) (with respect to its detailed versions (9)-(11)) contains all information we need. Namely, it determines the fields (electric and magnetic) and temperature dependences of order parameters *P*, *M* and *L* and hence their field (and temperature if necessary) derivatives also.

The set (9) – (11) has a plenty of solutions depending on the phase we choose. The experiment to be described suggests us what class of solutions we are to pick up. Namely, as we do not have experimental dependences on electric field, we put it to be zero: *E*=0. The second fact is that ferroelectric $T_C$=370 K is higher then $T_N$=150 K so that the spontaneous polarization *P* exists and we do not choose the solution *P*=0 (at *E*=0) of Eq. (9). Rather, we cancel down *P* in (9) at *E*=0 and obtain the following equation for spontaneous polarization *P=P(T,H)*

$$\alpha_P + \xi_{MP}M^2 + \xi_{LP}L^2 + \beta_P P^2 = 0. \tag{12}$$

The set of equations (12) and (10), (11) defines the necessary solutions for our experimental case of interest. Namely, in paramagnetic case we should consider solutions with *L*=0 while in AFM one we consider $L \neq 0$.

*1. Paramagnetic phase*

In paramagnetic phase we consider the solution *L*=0 so that Eq. (11) is satisfied identically. The set of remaining equations at *L*=0 assumes the form

$$\alpha_P + \xi_{MP} M^2 + \beta_P P^2 = 0, \tag{13}$$

$$M(\alpha_M + \xi_{MP} P^2) + \beta_M M^3 = H. \tag{14}$$

Note that while here $P$ is spontaneous (electric field independent) polarization, the magnetization $M$ is entirely generated by magnetic field. In principle, the nonlinear term $\beta_M M^3$ can be neglected but we will not do that as the set (13), (14) is very simple and admits an analytical expression for magnetoelectric coefficient. Namely, from (13)

$$P^2 = \frac{-\alpha_P - \xi_{MP} M^2}{\beta_P} \Rightarrow P = P_s(T)\sqrt{1 - \frac{\xi_{MP} M^2}{|\alpha_P(T)|}}, \tag{15}$$

where we explicitly show the temperature dependent quantities. Here $P_s^2 = -\alpha_P / \beta_P > 0$ is a spontaneous polarization value in the "pure ferroelectric" (i.e. without magnetic contribution) case, where $\alpha_P < 0$. The derivative $dP/dH$ can be easily obtained from (15)

$$U_{ME} \sim \frac{dP}{dH} = -\frac{P_s(T)\xi_{MP}}{|\alpha_P(T)|\sqrt{1 - \frac{\xi_{MP} M^2}{|\alpha_P(T)|}}} M \frac{dM}{dH} \equiv -\frac{\xi_{MP}}{\sqrt{\beta_P}\sqrt{|\alpha_P(T)| - \xi_{MP} M^2}} M \frac{dM}{dH}. \tag{16}$$

The last expression in (16) shows that the temperature dependence of $\beta \sim dP/dH$ in paramagnetic phase comes from $|\alpha_P(T)|$ under square root in the denominator and hence is weak. Other important fact is that we already have negative sign of $\beta$. This is because $\xi_{MP} > 0$ (at least this can be asserted from comparison with experiment) and $MM' = (1/2)(dM^2/dH) > 0$ as $M(H)$ is an increasing function.

The dependence $M(H)$ can be obtained from (10) by substitution of $P^2$ from Eq. (15), which yields

$$Q_1 M + Q_2 M^3 = H, Q_1 = \alpha_M(T) + \frac{|\alpha_P(T)|\xi_{MP}}{\beta_P}, Q_2 = \beta_M - \frac{\xi_{MP}^2}{\beta_P}. \tag{17}$$

In this case

$$\frac{dM}{dH} = \frac{1}{Q_1 + 3Q_2 M(H)^2}. \tag{18}$$

To find the dependence $U_{ME}(H,T) \sim dP/dH$ numerically, we first solve equation (17) at a given temperature for $M(H)$ and then substitute the obtained solution to Eq. (16) with respect to relation (18). Our numerical solution shows (see also below) that the dependence $U_{ME}(H,T)$ is a linear function of $H$ and in the experimentally available temperature range (~150 - 200 K) it is almost temperature independent.

To obtain $U_{ME}(H,T)$ analytically, we look for solution of Eq. (17) in the form

$$M = \chi_1 H + \chi_3 H^3 + \ldots + \chi_{2n+1} H^{2n+1}. \tag{19}$$

Substitution of this expression into Eq. (16) gives for the first two coefficients (we can calculate as much coefficients as possible) $\chi_1 = 1/Q_1$, $\chi_3 = -Q_2/Q_1^4$ so that

$$M \approx \frac{H}{Q_1} - \frac{Q_2}{Q_1^4} H^3. \tag{20}$$

Then, in the lowest approximation in $H$

$$P \approx P_s \left(1 - \frac{1}{2} \frac{\xi_{MP} H^2}{Q_1^2 |\alpha_P|}\right), \tag{21}$$

$$U_{ME} \sim \frac{dP}{dH} = -\frac{\xi_{MP}}{Q_1^2 \sqrt{|\alpha_P|\beta_P}} H. \tag{22}$$

The expression (22) is the solution in the paramagnetic phase. In accord with experiment it has the form of linear function of $H$ with negative slope. Below we will rewrite its temperature dependence explicitly and plot this dependence.

## *2. Antiferromagnetic phase*

Although this case is more complicated than the paramagnetic one, the method of solution here is essentially the same. In this case we consider the solution with $L \neq 0$ (spontaneous AFM moment) of the equation (11). Canceling $L$ in (11), we obtain now the following system of equations instead of (13), (14)

$$\alpha_L + \xi_{LM} M^2 + \xi_{LP} P^2 + \beta_L L^2 = 0. \tag{23}$$

$$M(\alpha_M + \xi_{LM} L^2 + \xi_{MP} P^2) + \beta_M M^3 = H. \tag{24}$$

Equations (12), (23), (24) constitute the "master" set of equations for the AFM phase. Equation (23) permits to express $L^2$ via $M^2$ and $P^2$ and eliminate it from (24). This yields

$$M\left[\alpha_M + \xi_{LM} L_0^2 + P^2 \left(\xi_{MP} - \frac{\xi_{LM}\xi_{LP}}{\beta_L}\right)\right] + M^3\left(\beta_M - \frac{\xi_{LM}^2}{\beta_L}\right) = H. \tag{25}$$

Here we introduce the equilibrium spontaneous AFM moment $L_0^2 = -\alpha_L/\beta_L$. The next step is to use the same trick to eliminate $L^2$ from Eq. (12) thus expressing $P^2$ via $M^2$ only. We have

$$P^2 = P_s^2 - \frac{\xi_{MP}}{\beta_P} M^2 - \frac{\xi_{LP}}{\beta_P} L^2 = P_s^2 - \frac{\xi_{MP}}{\beta_P} M^2 - \frac{\xi_{LP}}{\beta_P}\left(L_0^2 - \frac{\xi_{LM}}{\beta_L} M^2 - \frac{\xi_{LP}}{\beta_L} P^2\right) =$$

$$P_s^2 - \frac{\xi_{LP}}{\beta_P} L_0^2 - M^2\left(\frac{\xi_{MP}}{\beta_P} + \frac{\xi_{LM}}{\beta_L}\right) + \frac{\xi_{LP}^2}{\beta_P \beta_L} P^2. \tag{26}$$

The equation (26) for $P^2$ looks complicated but actually it has a simple form $P^2 = a_1 + a_2 P^2$. Its solution has the form $P^2 = a_1/(1-a_2)$ or explicitly

$$P^2 = A_0 + B_0 M^2, \quad A_0 = \frac{P_s^2 - \frac{\xi_{LP}}{\beta_P} L_0^2}{1 - \frac{\xi_{LP}^2}{\beta_P \beta_L}}, \quad B_0 = \frac{1}{\beta_P} \frac{\frac{\xi_{LM} \xi_{LP}}{\beta_L} - \xi_{MP}}{1 - \frac{\xi_{LP}^2}{\beta_P \beta_L}}. \tag{27}$$

Substitution of Eq. (27) into Eq. (25) generates the following equation for $M(H)$

$$U_1 M + U_2 M^3 = H, \tag{28}$$

$$U_1 = \alpha_M + \xi_{LM} L_0^2 + A_0 \left( \xi_{MP} - \frac{\xi_{LM} \xi_{LP}}{\beta_L} \right),$$

$$U_2 = \beta_M - \frac{\xi_{LM}^2}{\beta_L} + B_0 \left( \xi_{MP} - \frac{\xi_{LM} \xi_{LP}}{\beta_L} \right).$$

The equation (28) has a form similar to Eq. (17). To obtain coefficients $Q_1$ and $Q_2$, we should not only put $L_0=0$ in $U_1$ and $U_2$ but also put $\xi_{LM} = \xi_{LP} = 0$. The comparison of $P^2(M^2)$ dependences in the PM phase (15) and in the AFM phase (27) shows that while in the AFM phase $B_0>0$, in the PM phase ($\xi_{LM} = \xi_{LP} = 0$) $B_0$ becomes negative which is responsible for the sign change of magnetoelectric coefficient while traversing $T_N$.

Proceeding along the same lines, as in the PM phase, we obtain the desired dependence $\beta(T,H)$ in the AFM phase, suitable for numerical calculations

$$U_{ME} \sim \frac{dP}{dH} = \frac{B_0}{\sqrt{A_0 + B_0 M^2}} M \frac{dM}{dH}, \tag{29}$$

$$\frac{dM}{dH} = \frac{1}{U_1 + 3M^2 U_2}.$$

The numerical calculations with the help of Eq. (29) show that it is possible to obtain very good analytical approximation to this expression. Namely, looking for approximate solution of Eq. (28) in the form (19), we obtain

$$M \approx \frac{H}{U_1} - \frac{U_2}{U_1^4} H^3 \tag{30}$$

and further substitute this solution to (29). This gives the analytical formula, which numerical outcome is indistinguishable from that of (29)

$$\frac{dP}{dH} = \frac{B_0 H}{U_1^2} \frac{\dfrac{1}{U_1} - \dfrac{4U_2}{U_1^4} H^2}{\sqrt{A_0 + \dfrac{B_0}{U_1^2} H^2 - \dfrac{2B_0 U_2}{U_1^5} H^4}}. \tag{31}$$

The approximate analytical solution in the form of power series gives

$$\frac{dP}{dH} \approx \frac{B_0 H}{U_1^2 \sqrt{A_0}} \left[1 - H^2 \frac{U_1 B_0 + 8U_2 A_0}{2 A_0 U_1^3}\right], \quad \beta = \frac{1}{H}\frac{dP}{dH} \approx \frac{B_0}{U_1^2 \sqrt{A_0}} \left[1 - H^2 \frac{U_1 B_0 + 8U_2 A_0}{2 A_0 U_1^3}\right], \tag{32}$$

showing explicitly that the ME coefficient $\beta$ is positive for low fields but it decreases with field increase.

**D. Numerical results**

*1. Theoretical curves*

It is convenient to introduce the dimensionless temperature $\tau = T/T_C$ and coefficient $\kappa = T_C/T_N > 1$. As in PFN $T_C \approx 370$ K, $T_N \approx 150$ K and $\theta \approx -520$ K, in these units the ferroelectric phase is realized at $\tau < 1$ and AFM phase at $\tau < 1/\kappa = 15/37 \approx 0.405$. Hence PM phase occurs at $0.405 < \tau < 1$. Now we rewrite the temperature dependent coefficients in the above units to obtain

$$A_0 = \frac{\alpha_{P_0} T_C}{\beta_P \left[1 - \dfrac{\xi_{LP}^2}{\beta_P \beta_L}\right]} \left[1 - \tau - \kappa_2(1 - \kappa\tau)\right] \equiv a_1 + a_2 \tau, \tag{33}$$

$$a_1 = \frac{\alpha_{P_0} T_C}{\beta_P \left[1 - \dfrac{\xi_{LP}^2}{\beta_P \beta_L}\right]} (1 - \kappa_2), \quad a_2 = \frac{\alpha_{P_0} T_C}{\beta_P \left[1 - \dfrac{\xi_{LP}^2}{\beta_P \beta_L}\right]} (\kappa_2 \kappa - 1), \quad \kappa_2 = \frac{\xi_{LP} \alpha_{L_0} T_N}{\beta_L \alpha_{P_0} T_C}.$$

$$Q_1 = \alpha_{M_0} T_C \left(\tau + \frac{|\theta|}{T_C}\right) + \frac{\xi_{MP} \alpha_{P_0} T_C}{\beta_P}(1 - \tau) \equiv q_{11} + q_{12}\tau, \quad \sqrt{|\alpha_P|} = q_{13}\sqrt{1 - \tau}, \tag{34}$$

$$q_{11} = \alpha_{M_0}|\theta| + \frac{\xi_{MP} \alpha_{P_0} T_C}{\beta_P}, \quad q_{12} = \alpha_{M_0} T_C - \frac{\xi_{MP} \alpha_{P_0} T_C}{\beta_P}, \quad q_{13} = \sqrt{\alpha_{P_0} T_C}.$$

$$U_1 = g_1\left(\tau + \frac{|\theta|}{T_C}\right) + g_2(1 - \kappa\tau) + g_3\left[1 - \tau - \kappa_2(1 - \kappa\tau)\right] \equiv u_{11} + u_{12}\tau, \tag{35}$$

$$g_1 = \alpha_{M_0} T_C, \quad g_2 = \frac{\xi_{LM} \alpha_{L_0} T_N}{\beta_L}, \quad g_3 = \frac{\alpha_{P_0} T_C}{\beta_P \left[1 - \dfrac{\xi_{LP}^2}{\beta_P \beta_L}\right]} \left(\xi_{MP} - \frac{\xi_{LM} \xi_{LP}}{\beta_L}\right),$$

$$u_{11} = g_1 \frac{|\theta|}{T_C} + g_2 + g_3(1 - \kappa_2), \quad u_{12} = g_1 - g_2\kappa + g_3(\kappa_2\kappa - 1).$$

The reduced temperatures equal to $|\theta|/T_C \equiv 520/370 = 1.405$, $\kappa = T_C/T_N = 2.47$. Having temperature dependent parameters (33)-(35), we can write the dependences $U_{ME}(H,\tau)$ in both PM and AFM phases explicitly

$$U_{ME} \sim \frac{dP}{dH} = -p_0 \frac{H}{(q_{11}+q_{12}\tau)^2 \sqrt{1-\tau}}, \quad p_0 = \frac{\xi_{MP}}{q_{13}\sqrt{\beta_P}}, \text{ PM phase}, \quad (36)$$

$$U_{ME} \sim \frac{B_0 H}{(u_{11}+u_{12}\tau)^{5/2}} \frac{(u_{11}+u_{12}\tau)^3 - 4U_2 H^2}{\sqrt{(a_1+a_2\tau)(u_{11}+u_{12}\tau)^5 + B_0(u_{11}+u_{12}\tau)^3 H^2 - 2B_0 U_2 H^4}}, \text{ AFM phase} \quad (37)$$

where $B_0$ and $U_2$ are also (fitting) parameters. Note that magnetic field $H$ here can be regarded both as dimensional and dimensionless quantity. Really, if we divide $H$ by some $H_0$ and introduce dimensionless field $h = H/H_0$, we simply renormalize coefficients like $p_0$.

The field dependences (36), (37) are reported in Fig. 7. The close resemblance to corresponding experimental curves can be seen. Below we will fit these curves to real experimental data.

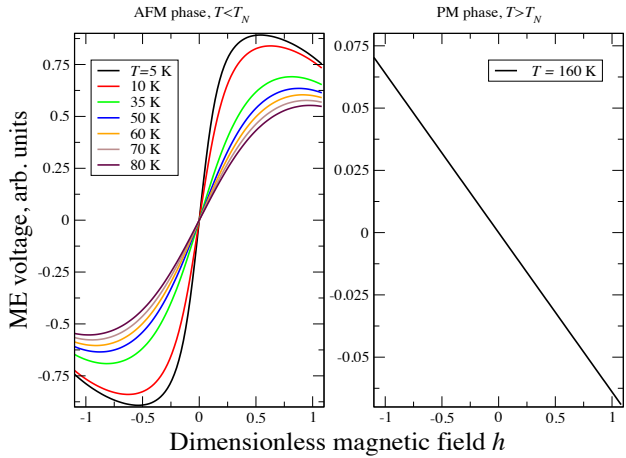

Fig. 7. Magnetic field dependence of ME voltage (arbitrary units) in AFM (left panel) and PM (right panel) phases at different temperatures, shown in the legend. The curves have been plotted using Eqs. (36), (37) with the following parameters: $B_0$=0.6, $U_2$=0.04, $u_{11}$=0.762162, $u_{12}$=0.376, $a_1$=0.01, $a_2$=2.9333.

We plot the temperature dependence of ME voltage in Fig. 8 at dimensionless magnetic field $h$=1 and other parameters being similar to those in Fig. 7. At $T=T_N$ the discontinuity in the temperature dependence is seen. This discontinuity is due to the fact that coefficients $\xi_{LM}$ and $\xi_{LP}$ do not have temperature dependence and appear abruptly in the AFM phase. In our opinion, the effects of weak site disorder in the Fe spins of PFN would, in accordance with experiment, lift this discontinuity. These effects give spin glass features like the difference in field cooled and field heated regimes, which are present in the vicinity of $T_N$ in the experimental curves. The smearing of the discontinuity may appear in the form of additional temperature dependence of the coefficients $\xi_{LM}$ and $\xi_{LP}$ so that they will no more appear abruptly in AFM phase.

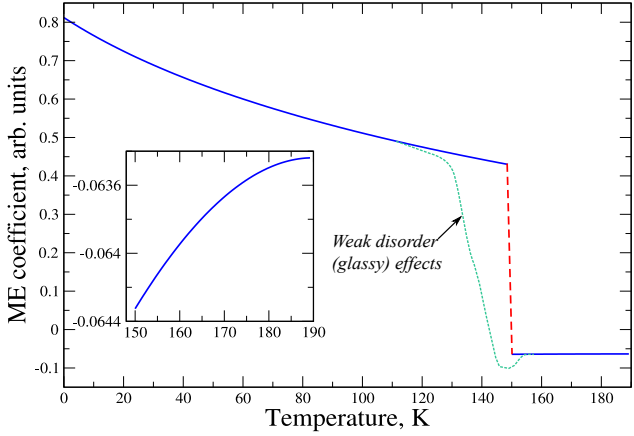

Fig. 8. Temperature dependence of ME coefficient (arbitrary units) at dimensionless magnetic field $h$=1. The parameters are similar to those from Fig. 8. At $T=T_N$ the discontinuity in the temperature dependence is seen. This discontinuity is attributed to the fact that the coefficients $\xi_{LM}$ and $\xi_{LP}$ do not have temperature dependence and appear abruptly in the AFM phase. The effects of weak disorder in the Fe spins in PFN lead to spin glass effects which smear the discontinuity at $T=T_N$. The inset highlights the temperature dependence in the PM phase at $T>T_N$=150 K.

## *2. Fitting to experiment*

The fitting of field dependences of ME voltage by the expressions (33) (PM phase) and (34) (AFM phase) is shown in Fig. 2 by solid lines. The best fit in the AFM phase (Fig. 2a) has been achieved for the following parameters values: $B_0$=1000, $U_2$=0.04, $u_{11}$=0.9, $u_{12}$=0.376, $a_1$=1.0, $a_2$=2.9933. For Fig. 2b (PM phase) is the best fit is achieved at $q_{11}$=3.0, $q_{12}$=2.2, $p_0$=185.85. We note that the latter sets of best fit parameters are not unique so that additional experiments may be required to determine unambiguously the coefficients of phenomenological free energy function (6) and hence the above parameters. Nevertheless, the excellent coincidence with experimental data is well seen. In our view this is because the simple phenomenological approach, although having many unknown coefficients, accounts correctly for order parameters coupling with magnetic (and electric) fields.

At the same time, visual comparison of theoretical curve $U_{ME}(T)$ at fixed $H$ in both phases and experimental ones shows that it is barely possible to achieve the quantitative coincidence between theory and experiment in this case. Our analysis confirms this point as we were not able to find suitable set of parameters which permits to achieve the slope of the curves $U_{ME}(T)$ observed in experiment at low temperatures. This is because the phenomenological LGD approaches usually have simplified temperature dependences of their coefficients, which do not account, e.g., for possible spin glass effects. Hence, although it is possible to achieve excellent quantitative coincidence in the $U_{ME}$ field dependences, the temperature dependences can be described only qualitatively. For quantitative description of temperature dependence, more sophisticated approaches, considering (weak) disorder effects, should be utilized.

## V. Discussion

We determined the quadratic ME coefficients for PFN and its solid solutions with PT and listed them in Table I. This table compares our values with the literature data for some other magnetoelectric materials. In particular, the ME coefficient is approximately the same in the PM phase of PFN and PFN-PT as in paramagnetic $Gd_2(MoO_4)_3$ and $NiSO_4 \cdot 6H_2O$ single crystals. In the AFM phase of PFN and PFN-PT, it increases by almost two orders of magnitude exhibiting sign reversal at the AFM phase transition. ME coefficient values for PFN and PFN-PT are by two-three orders of magnitude higher than those for the well-known multiferroic $BiFeO_3$ and magnetoelectrics $CsCuCl_3$, $Ni_3B_7O_{13}Cl$ and $BaMnF_4$ with weak ferroelectricity. Note that although ME effect is weak in $ScCuCl_3$, it shows the sign reversal at the AFM phase transition similar to PFN [25].

Table I. Quadratic $\beta_{333}$ ME coefficients in paramagnetic, antiferromagnetic and spin glass phases of PFN and PFN–PT solid solutions. SC stands for single crystal. For comparison, the measured quadratic ME coefficient is reported for other AFM magnetoelectrics.

| Material | PM phase | AFM phase | SG phase | References |
| --- | --- | --- | --- | --- |
| PFN, SC |  | $2.5 \cdot 10^{-17}$ s/A |  | this paper |
| PFN, SC |  | $(1-10) \cdot 10^{-17}$ s/A |  | [11,1] |
| PFN, ceramics | $-1 \cdot 10^{-18}$ s/A | $9.6 \cdot 10^{-17}$ s/A |  | this paper |
| PFN-0.03PT, SC |  | $4.8 \cdot 10^{-17}$ s/A |  | this paper |
| PFN-0.05PT ceramics | $-2 \cdot 10^{-18}$ s/A |  | $-1.5 \cdot 10^{-17}$ s/A | this paper |
| PFN-0.2PT, SC |  |  | $-1.45 \cdot 10^{-17}$ s/A | this paper |
| $BiFeO_3$, SC |  | $2.1 \cdot 10^{-19}$ s/A |  | [12,13] |
| $Gd_2(MoO_4)_3$, SC | $0.8 \cdot 10^{-18}$ s/A |  |  | [24] |
| NiSO$_4$·6H$_2$O, SC | 2.2·10$^{-9}$ esu<br>0.7·10$^{-18}$ s/A | | | [19] |
| CsCuCl$_3$, SC | -0.02 pC/(cm T)$^2$<br>-3.1·10$^{-22}$ s/A | 0.14 pC/(cm T)$^2$<br>2.2·10$^{-21}$ s/A | | [25] |
| Ni$_3$B$_7$O$_{13}$Cl, SC | | $|0.6\text{-}1.6|$·10$^{-18}$ s/A | | [26] |
| BaMnF$_4$, SC | | (0.8-1.6)·10$^{-19}$ s/A | | [27] |

Another important fact to which we want to draw attention is the strong nonlinearity of the ME effect in the AFM phase of PFN and PFN-PT even at low fields, 5 – 15 kOe. Usually nonlinearity of the ME effect appears in the vicinity of spin-flip, spin-flop or other magnetic field induced phase transitions. The phenomenological theory introduced in Section IV allows one to describe this nonlinearity by choosing appropriate values for coefficients of the free energy expansion. Obviously, the most important is the coefficient $\xi_{LP}$, which describes the coupling of ferroelectric polarization to AFM order parameter. This coefficient cannot be directly extracted from our experiment. However, it has to be much larger than that in other magnetoelectrics listed in Table I.

In spite of the fact that the ME effect is highly anisotropic (the quadratic ME coefficient is described by a third rank tensor; to the best of our knowledge it was measured in single crystals only) and should be averaged in ceramics, we found approximately the same values of the ME coupling coefficient in both single crystal and ceramic samples. This means that the magnetic anisotropy generates strong coupling of magnetic domain structure to electric polarization and ferroelectric domains. In other words, for instance, AFM domains can be aligned by external electric field similar to ferroelectric domains. This had been demonstrated previously for BiFeO$_3$ crystals [14]. Therefore, from the application point of view, PFN-based multiferroics are attractive in ME memory elements and spintronics as AFM domains are very stable with respect to external magnetic fields.

Fig. 9a illustrates the variation of ME response angular dependence in AFM phase of PFN ceramics after electric field polarity reversal with respect to the initial polarization direction. One can see that ME signal changes sign at electric field polarity reversal indicating 180$^0$ switching of polarization. Both curves are well fitted by the function

$$U_{ME} = A + B\cos(2\theta - \varphi), \tag{38}$$

where $\theta$ is the angle between electric polarization and the magnetic field and $\varphi=0$ or $180^0$ determines the electric field polarity. This demonstrates that the ME coefficient (proportional to $B\cos\varphi$) changes the sign at the electric field polarity reversal.

However, the above experiment does not permit to discern the influence of the above polarization switching on AFM order parameter $L$. This is because ME response is proportional to $L^2$ and both lattice distortion and magnetic anisotropy remain to be along the same crystal direction. Here we assume that magnetization and polarization are parallel to each other in accordance with neutron diffraction data [28] and ME measurements in single crystals [9,11]. However, if the polarization is switched by $71^0$ from the initial rhombohedral direction into equivalent one (in ceramics, the average polarization switches by $90^0$), the AFM order parameter should also change its orientation. This occurs due to the influence of magnetic anisotropy, which forces the magnetization to rotate along polarization direction in order to minimize the lattice energy with respect to a new rhombohedral distortion. Such $90^0$ switching of the polarization in the AFM phase of ceramic sample is demonstrated in Fig. 9b, which shows the change of ME voltage angular dependence under the electric field application perpendicular to initial polarization direction.

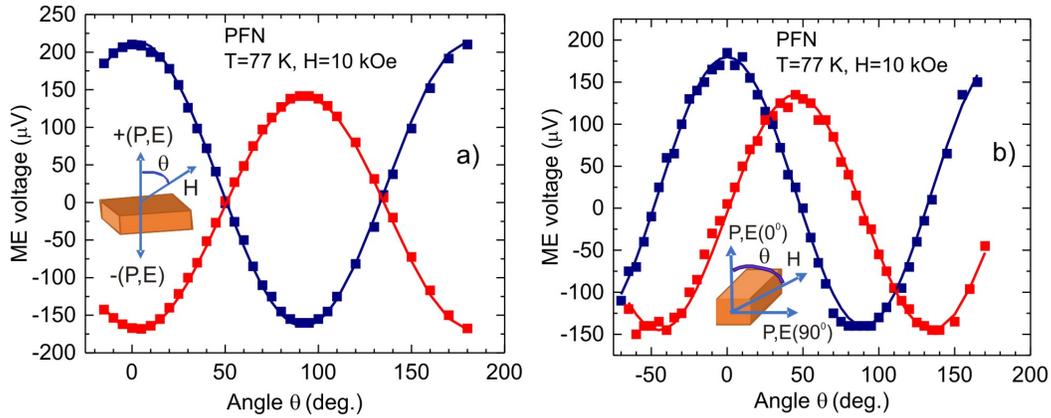

Fig. 9. Angular dependencies of ME voltage in PFN ceramics under the electric field polarity reversal (a) and $90^0$ switching of polarization (b) in the AFM phase at T=77 K and H=10 kOe. Solid lines are fit to Eq. (38).

One can see that two curves having approximately the same amplitude are now shifted from each other by the angle $\varphi=90^0$, which is two times smaller than in the case of $180^0$ switching of polarization. These data suggest that AFM order parameter $L$ also rotates parallel to electric polarization in the case of $90^0$ switching. In the opposite situation, when the magnetic order parameter retains initial direction, two curves in Fig. 10b should be essentially different as $P$ and $L$ are coupled by different ME coefficients corresponding to $P\|L$ or $P\perp L$. Certainly, in the course of sample cooling from PM to AFM phase under poling field, magnetic domains become aligned. Our results show that both ferroelectric and AFM domains can be directly switched by an electric field also in the AFM phase though the coercive field increases by almost ten times ($E_c \approx 25$ kV/cm) as compared to PM phase [7,10]. To confirm our latter conclusion, additional neutron diffraction or AFM resonance measurements can be used to monitor the AFM domains rotation. However, the fact that even in ceramics we detect approximately the same strong ME signal at different electric polarization directions suggests that electric domains alignment really causes corresponding adjustment of magnetic domains and this process can be controlled by electric field in the AFM phase.

### VI. Conclusions

**To summarize**, we have measured ME effect in PFN and PFN-PT solid solutions multiferroics. We perform our measurements both on crystalline and ceramic samples in a broad temperature range from liquid helium up to the room temperature. Surprisingly, the ME effect turns out to be approximately the same or even higher in ceramics than that in single crystals. In our opinion this is probably due to better possibility to pole the ceramics as its resistivity is much higher. This suggests that the alignment of polar domains leads to corresponding adjustment of magnetic domains as the magnetization axis is coupled to polarization direction via magnetic anisotropy. Latter anisotropy is high in PFN due to strong lattice distortions generated by ferroelectric phase transition. We also demonstrate the possibility of AFM domains switching by external electric field even in ceramic PFN samples. The latter effect could be attractive for spintronic and ME random access memory applications as ME coupling offers an elegant way of electric filed control and switching of AFM domains. It is worth noting that though the Neel temperature of PFN ($\approx 150$ K) is well below the room temperature, it can be increased substantially by mechanical activation of precursors used for ceramics sintering [29] and by strain engineering in epitaxial nanofilms [30,31].

There is a sign reversal of ME coefficient at the paramagnetic to antiferromagnetic phase transition. It indicates that the ME response related to AFM order parameter has the opposite sign to that in the paramagnetic phase. The PM-like contribution is non-zero in the AFM phase in spite of the fact that $Fe^{3+}$ spins are antiferromagnetically ordered. This contribution increases on cooling proportionally to increase of FC susceptibility. This fact strongly supports the model of the PFN ground magnetic state as coexistence of the long-range ordered AFM phase with the short-range ordered SG state on microscopic level (see, e.g. Refs. [3,8]). On the other hand, the ME coefficient does not show the sign reversal at the transition from the paramagnetic phase to the spin glass state. The ME coefficient is negative in both PM and SG states. Note that ME coefficient in PFN and PFN–PT is almost three orders of magnitude larger than that in $BiFeO_3$. It is obviously related to the fact that $L^2P^2$ ME coupling is essentially averaged by spin rotations along $BiFeO_3$ cycloid so that $M^2P^2$ term (Eqs. (3) and (6)) contributes primarily to ME response similar to PM-FE phase of PFN.

While the ME response is a perfect linear function of applied magnetic field in PM or SG phases, it becomes strongly nonlinear in the AFM phase even at low magnetic fields of only few kOe. We naturally explain this phenomenon in the framework of Landau theory by choosing appropriate values for coefficients of the free energy expansion.

**Acknowledgments**

The research was supported by the GA CR under project No. 13-11473S, projects SAFMAT LM2015088 and LO1409.

[1] W. Kleemann, V. V. Shvartsman, P. Borisov, A. Kania, *Phys. Rev. Lett*. **2010**, *105*, 257202.

[2] V. V. Laguta, M. D. Glinchuk, M. Maryško, R. O. Kuzian, S. A. Prosandeev, S. I. Raevskaya, V. G. Smotrakov, V. V. Eremkin, I. P. Raevski, *Phys Rev B* **2013**, 87, 064403.

[3] S. Chillal, M. Thede, F. J. Litterst, S. N. Gvasaliya, T. A. Shaplygina, S. G. Lushnikov, A. Zheludev, *Phys. Rev. B* **2013**, *87*, 220403(R).

[4] S. P. Singh, S. M. Yusuf, S. Yoon, S. Baik, N. Shin, D. Pandey, *Acta Mat.* **2010**, *58*, 5381-5392.

[5] D. A. Sanches, N. Ortega, A. Kumar, G. Sreenivasulu, R. S. Katiyar, J. F. Scott, D. M. Evans, M. Arredondo-Arechavala, A. Schilling, J. M. Gregg, *J. Appl. Phys*. **2013**, *113*, 074105.

[6] I. P. Raevski, S. P. Kubrin, S. I. Raevskaya, S. A. Prosandeev, M. A. Malitskaya, V. V, Titov, D. A. Sarychev, A. V. Blazhevich, I. N. Zakharchenko, *IEEE Trans. Ultrason. Ferroelect. Freq. Contr.* **2012**, *59*, 1872-1878.

[7] E. I. Sitalo, I. P. Raevski, A. G. Lutokhin, A. V. Blazhevich, S. P. Kubrin, S. I. Raevskaya, Y. N. Zakharov, M. A. Malitskaya, V. V. Titov, I. N. Zakharchenko, (2011) *IEEE Trans. Ultrason. Ferroelect. Freq. Contr.* **2011**, *58*, 1914-1919.

[8] V.A. Stephanovich, V.V. Laguta, *Phys. Chem. Chem. Phys.* **2016**, *18*, 7229-7234.

[9] T. Watanabe, K. Kohn, *Phase Transitions* 1989, *15*, 57-68.

[10] V.V. Laguta, A. N. Morozovska, E. A. Eliseev, I. P. Raevski, S. I. Raevskaya, E. I. Sitalo, S. A. Prosandeev, L. Bellaiche, *J. Mater. Science* **2016**, *51*, 5330-5342.

[11] B. Howes, M. Pelizzone, P. Fischer, C. Tabaresmunoz, J.-P. Rivera, H. Schmid, (1984) *Ferroelectrics* 1984, *54*, 317-320.

[12] C. Tabares-Munoz, J.-P. Rivera, A. Bezinges, A. Monnier, H. Schmid, (1985) *Jpn. J. Appl. Phys*. 1985, *24* (part 1), 1051-1053.

[13] V. A. Murashov, D. N. Rakov, N. A. Ekonomov, A. K. Zvezdin, I. S. Dubenko, *Solid State Physics* **1990**, *32*, 2156-2158.

[14] D. Lebeugle, D. Colson, A. Forget, M. Viret, A. M. Bataille, and A. Gukasov, *Phys. Rev. Lett*. 2008, *100*, 227602.

[15] D. Rahmedov, S. Prosandeev, J. Iniguez, L. Bellaiche, (2013) *Phys. Rev. B* **2013**, *88*, 224405.

[16] E. A. Dul'kin, I. P. Raevskii, S. M. Emel'yanov, Phys. Solid State **1997**, *39*, 316-317.

[17] I. P. Raevskii, S. T. Kirillov, M. A. Malitskaya, V. P. Filippenko, S. M. Zaitsev, L. G. Kolomin, *Inorg. Mate*r. **1988**, *24*, 217-220.

[18] M. M. Kumar, A. Srinivas, S. V. Surynarayana, G. S. Kumar, T. Bhimasankaram, (1998). *Bull. Mater. Sci.* **1998**, *21*, 251-255 A

[19] S. L. Hou, N. Bloembergen, *Phys. Rev*. **1965**, *138*, A1218-A1226.

[20] M. Fiebig, *J. Phys. D: Appl. Phys*. **2005**, *38*, R123-R152.

[21] T. Murao, *Prog. Theor. Phys*. **1967**, *37*, 1038-1040.

[22] I. P. Raevski, S. P. Kubrin, S. I. Raevskaya, V. V. Stashenko, D. A. Sarychev, M. A. Malitskaya, M. A. Seredkina, V. G. Smotrakov, I. N. Zakharchenko, V. V. Eremkin, *Ferroelectrics* **2008**, *373*, 121-126.

[23] R. O. Kuzian, I. V. Kondakova, A. M. Dar´e, V. V. Laguta, *Phys. Rev. B* **2014**, *89*, 024402.

[24] B. K. Ponomarev, E. Stiep, H. Wiegelmann, A.G.M. Jansen, W. Wyder, B. S. Red'kin, *Phys. Solid State* **2000**, *42*, 734.

[25] A. I. Kharkovskiy, Yu. V. Shaldin, V. I. Nizhankovskii, *J. Appl. Phys*. **2016**, *119*, 014101.

[26] J.-P. Rivera and H. Schmid, *J. Appl. Phys*. **1991**, *70*, 6410-6412.

[27] A. K. Zvezdin, G. P. Vorob'ev, A. M. Kadomsteva, Yu. F. Popov, D. V. Belov, A. P. Pyatakov, *J. Exp. Theor. Phys*. **2009**, *109*, 221-226.

[28] S. Ivanov, R. Tellgren, H. Rundlof, N. W. Thomas, S. Ananta, *J. Phys.: Condens. Matter* **2000**, *12*, 2393.

[29] A. A. Gusev, S. I. Raevskaya, V. V. Titov, V. P. Isupov, E. G. Avvakumov, I. P. Raevski, H. Chen, C.-C. Chou, S. P. Kubrin, S. V. Titov, M. A. Malitskaya, D. A. Sarychev, V. V. Stashenko, S. I. Shevtsova, *Ferroelectrics* **2016**, *496*, 231-239.

[30] W. Peng, N. Lemee, M. Karkut, B. Dkhil, V. Shvartsman, P. Borisov, W. Kleemann, J. Holc, M. Kosec, R. Blinc, *Appl. Phys. Lett*. **2009**, *94***,** 012509.

[31] S. A. Prosandeev, I. P. Raevski, S. I. Raevskaya, H. Chen, *Phys. Rev. B*. **2015**, *92*, 220419(R).